\documentclass[%
 aip,
 amsmath,amssymb,
 reprint,%
]{revtex4-1}

\usepackage{graphicx}
\usepackage{dcolumn}
\usepackage{bm}

\usepackage[utf8]{inputenc}
\usepackage[T1]{fontenc}
\usepackage{mathptmx}
\usepackage{etoolbox}

\usepackage{color} 
\usepackage{ulem} 
\usepackage{siunitx}

\makeatletter
\def\@email#1#2{%
 \endgroup
 \patchcmd{\titleblock@produce}
  {\frontmatter@RRAPformat}
  {\frontmatter@RRAPformat{\produce@RRAP{*#1\href{mailto:#2}{#2}}}\frontmatter@RRAPformat}
  {}{}
}%
\makeatother
\begin{document}

\preprint{AIP/123-QED}

\title{Presence of a chiral soliton lattice in the chiral helimagnet MnTa$_{3}$S$_{6}$}

\author{Y. Miyagi}
\affiliation{Department of Physics and Electronics, Osaka Metropolitan University, Sakai, Osaka 599-8531 Japan}

\author{J. Jiang}
\affiliation{Department of Physics and Electronics, Osaka Metropolitan University, Sakai, Osaka 599-8531 Japan}

\author{K. Ohishi}
\affiliation{Neutron Science and Technology Center, Comprehensive Research Organization for Science and Society (CROSS), Tokai, Ibaraki 319-1106, Japan}

\author{Y. Kawamura}
\affiliation{Neutron Science and Technology Center, Comprehensive Research Organization for Science and Society (CROSS), Tokai, Ibaraki 319-1106, Japan}

\author{J. Suzuki}
\affiliation{Neutron Science and Technology Center, Comprehensive Research Organization for Science and Society (CROSS), Tokai, Ibaraki 319-1106, Japan}

\author{D. R. Alshalawi}
\affiliation{Neutron Science and Technology Center, Comprehensive Research Organization for Science and Society (CROSS), Tokai, Ibaraki 319-1106, Japan}
\affiliation{Institute of Applied Magnetism, UCM-ADIF-CSIC, Las Rozas 28230, Spain}
\affiliation{Department of Materials Physics, Complutense University of Madrid, Madrid 28040, Spain}

\author{J. Campo}
\affiliation{Arag\'{o}n Nanoscience and Materials Institute (CSIC -- University of Zaragoza) and Condensed Matter Physics Department, C/Pedro Cerbuna 12, 50009 Zaragoza, Spain}

\author{Y. Kousaka*}
\affiliation{Department of Physics and Electronics, Osaka Metropolitan University, Sakai, Osaka 599-8531 Japan}
\email{koyu@omu.ac.jp}

\author{Y. Togawa}
\affiliation{Department of Physics and Electronics, Osaka Metropolitan University, Sakai, Osaka 599-8531 Japan}

\date{\today}

\begin{abstract}
Chiral helimagnetism was investigated in transition-metal intercalated dichalcogenide single crystals of MnTa$_3$S$_6$. Small-angle neutron scattering (SANS) experiments revealed the presence of harmonic chiral helimagnetic order, which was successfully detected as a pair of satellite peaks in the SANS pattern. The magnetization data are also supportive of the presence of chiral soliton lattice (CSL) phase in MnTa$_3$S$_6$. The observed features are summarized in the phase diagram of MnTa$_3$S$_6$, which is in strong contrast with that observed in other dichalcogenides such as CrNb$_3$S$_6$ and CrTa$_3$S$_6$. The presence of the remanent state provides tunable capability of the number of chiral solitons at zero magnetic field in the CSL system, which may be useful for memory device applications. 
 
\end{abstract}

\maketitle



\section{Introduction}

\begin{figure}[tb]
\begin{center}
\includegraphics[width=8.5cm]{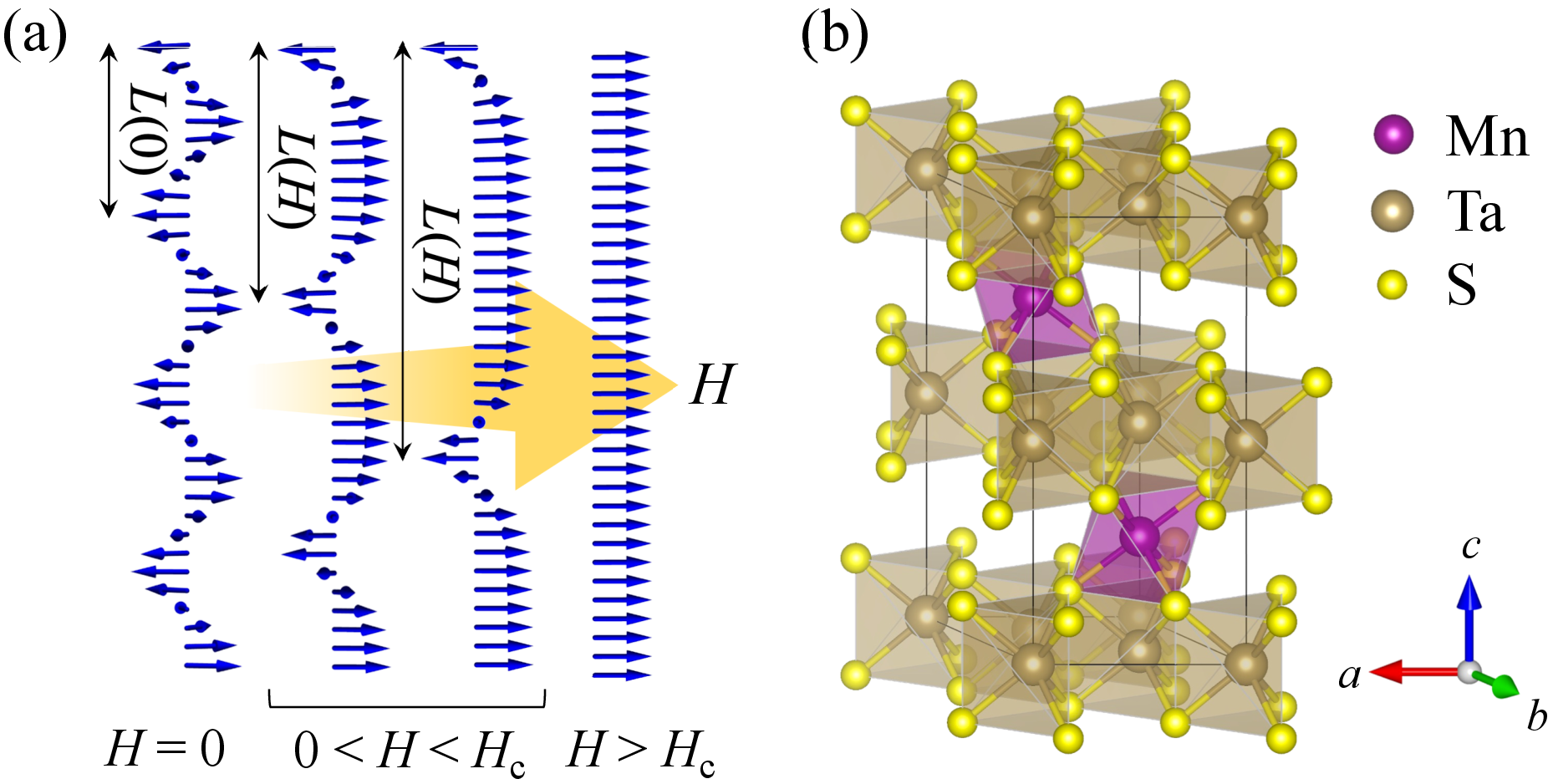}
\end{center}
\caption{
(a) Evolution of chiral magnetic structures when increasing the magnetic field perpendicular to the $c$-axis.
The chiral helimagnetic order (CHM) transforms into a forced ferromagnetic state (FFM) via a chiral soliton lattice (CSL).
(b) Crystal structure of MnTa$_{3}$S$_{6}$.
}
\label{f-str}
\end{figure}

Chiral helimagnets generate chiral magnetic structures because of a competition between antisymmetric Dzyaloshinskii-Moriya (DM) interaction~\cite{Dzyaloshinskii1958, Moriya1960} and symmetric Heisenberg exchange interaction. 
A chiral helimagnetic structure (CHM) appears at zero magnetic field~\cite{Moriya1982, Miyadai1983}, while a nontrivial helicoidal transformation from the CHM to a chiral soliton lattice (CSL)~\cite{Dzyaloshinskii1964, Dzyaloshinskii1965a, Dzyaloshinskii1965b, Izyumov1984, Kishine2005, Togawa2012,Shinozaki2016,Laliena2016-1,Laliena2016-2,Laliena2017,Shinozaki2018,Masaki2018} occurs in the presence of magnetic fields, as schematically drawn in Fig.~\ref{f-str}(a). 
Chiral magnetic vortices called magnetic Skyrmions also appear in some chiral helimagnets~\cite{Bogdanov1989, Bogdanov1994, Muhlbauer2009, Yu2010}.

The CSL is a tunable superlattice of magnetic moments with robust phase coherence~\cite{Togawa2012} and thus topological and collective responses appear in the CSL system. For instance, the CSL exhibits multivalued MR effect~\cite{Togawa2015, Tang2018}, nonreciprocal electrical transport~\cite{Aoki2019} and collective elementary excitation in a sub-terahertz frequency regime~\cite{Shimamoto2022}.
Moreover, magnetic Goos H\"{a}nchen effect  was demonstrated in uniaxial helimagnets with CSL~\cite{Laliena2022}.
Such unique features may open up a novel route of spintronic, magnonic and 6G communication device applications using the CSL~\cite{Kishine2015, Togawa2016, Togawa2023}. 

Transition-metal intercalated dichalcogenides (TMD) $M'M_{3}$S$_{6}$ ($M'$ = $3d$ transition metal,  $M$ = Nb or Ta) has attracted attention since the observation of the CHM and CSL in CrNb$_3$S$_6$.  It forms a chiral monoaxial crystal structure with a space group $P6_{3}22$~\cite{Berg1968, Anzenhofer1970, Laar1971}, as shown in Fig.~\ref{f-str}(b). 
In CrNb$_3$S$_6$, the CHM formation was detected by small-angle neutron scattering (SANS) experiments~\cite{Miyadai1983}, while the CSL formation was directly observed by Lorentz microscopy~\cite{Togawa2012}.

Note that the CHM period at zero magnetic field $L(0)$ is given as $2\pi (J/D) a_{\rm 0}$, where $J$ is the symmetric exchange interaction,
$D$ is that of DM interaction and $a_{\rm 0}$ is the size of unit cell along the $c$-axis.
A typical period of the CHM is several tens of nm because $D$ is much smaller than that of $J$. 
The corresponding wave number of the harmonic CHM is typically 0.1 nm$^{-1}$ or less.
Thus, neutron scattering experiments should be performed with a spatial resolution in $Q$ enough to probe magnetic satellite peaks~\cite{Moriya1982, Miyadai1983}.
The magnetic structures of TDMs were sometimes misinterpreted as ferromagnetic (FM) because of the difficulty to discern these magnetic satellite peaks by using SANS experiments.

Moreover, it is important to grow high-quality TMD crystals that host the CHM and CSL because a small amount of defects of intercalated transition metals stabilizes the FM ordering rather than the CHM and CSL formation~\cite{Kousaka2022}.
Indeed, MnNb$_3$S$_6$ and CrTa$_3$S$_6$ were first reported to exhibit the ferromagnetic order~\cite{Berg1968, Laar1971, Hulliger1970, Friend1977, Parkin1980a,Parkin1980b}.   However, in fact, high quality single crystals of these compounds were reexamined and found to form the CHM and CSL~\cite{Kousaka2009, Togawa2013, Kousaka2016, Zhang2021, Obeysekera2021,Mizutani2023}.
The values of a critical temperature $T_{\rm c}$, a critical magnetic field $H_{\rm c}$, $L(0)$ and $J/D$ in the chiral helimagnetic TMDs are summarized in Table~\ref{tbl-params}.

\begin{table}[htbp]
      \caption{
      A list of $T_{\rm c}$, $H_{\rm c}$, the CHM period $L(0)$, and $J/D$ for chiral helimagnetic TMDs.
      The values of the Cr-intercalated TMDs are taken from the literature for CrNb$_3$S$_6$~\cite{Miyadai1983,Kousaka2009,Kousaka2022} and CrTa$_3$S$_6$~\cite{Kousaka2016,Zhang2021,Obeysekera2021,Mizutani2023}.
      The values of MnTa$_3$S$_6$ are deducted from this work.
      }
      \label{tbl-params}

  \begin{center}

      \begin{tabular}{rrrrrrrr}
      \hline
      \multicolumn{1}{c}{} & \multicolumn{1}{c}{$T_{\rm c}$ (K)} & \multicolumn{1}{c}{} & \multicolumn{1}{c}{$H_{\rm c}$ (T)} & \multicolumn{1}{c}{} & \multicolumn{1}{c}{$L(0)$ (nm)} & \multicolumn{1}{c}{} & \multicolumn{1}{c}{$D/J$}\\
      \hline
      \multicolumn{1}{c}{CrNb$_{3}$S$_{6}$} & \multicolumn{1}{c}{133} & \multicolumn{1}{c}{} & \multicolumn{1}{c}{0.2} & \multicolumn{1}{c}{} & \multicolumn{1}{c}{48} & \multicolumn{1}{c}{} & \multicolumn{1}{c}{0.16}\\
      \multicolumn{1}{c}{CrTa$_{3}$S$_{6}$} & \multicolumn{1}{c}{150} & \multicolumn{1}{c}{} & \multicolumn{1}{c}{1.7} & \multicolumn{1}{c}{} & \multicolumn{1}{c}{22} & \multicolumn{1}{c}{} & \multicolumn{1}{c}{0.35}\\
      \multicolumn{1}{c}{MnTa$_{3}$S$_{6}$} & \multicolumn{1}{c}{35} & \multicolumn{1}{c}{} & \multicolumn{1}{c}{0.7} & \multicolumn{1}{c}{} & \multicolumn{1}{c}{86} & \multicolumn{1}{c}{} & \multicolumn{1}{c}{0.09}\\
      \hline
      \end{tabular}

  \end {center}
\end{table}

It was reported that MnTa$_3$S$_6$, which forms the same crystal structure as CrNb$_3$S$_6$,
showed ferromagnetism with a Curie temperature $T_{\rm{C}}$ of 70 K with the magnetic easy axis lying in the hexagonal basal plane \cite{Berg1968, Laar1971, Hulliger1970, Parkin1980a, Parkin1980b, Zhang2018, Algarni2023}.  When an external magnetic field $H$ was applied in the direction perpendicular to the $c$-axis, the magnetization showed a small anomaly at 35 K.   The magnetization curve at 2 K presented a tiny hysteresis loop near the saturation field \cite{Zhang2018}.  Nanoflake samples exhibited a topological Hall effect in $H$ parallel to the $c$-axis~ \cite{Algarni2023}.
However, no studies were reported about the presence of the CHM and CSL formation in MnTa$_3$S$_6$ single crystals, which constitute the main focus of this paper.

We investigated the magnetic structures of MnTa$_3$S$_6$ using SANS and magnetization experiments with high-quality single crystals.
The SANS experiments detected magnetic satellite peaks along the $c$-axis, indicating the CHM with the period of \SI{86}{\nano \metre}.
The magnetization measurements, via a zero-field cooling process, also revealed the CSL formation as observed in other TMDs such as CrNb$_3$S$_6$ and CrTa$_3$S$_6$.  The hysteresis in the magnetization curves makes a magnetic phase diagram complicated as compared to other TMDs that exhibit chiral helimagnetism.
However, the presence of the remanent state may open up the development of memory device applications using chiral solitons as multiple information bits.

\section{Experimental Methods}

Single crystals of MnTa$_{3}$S$_{6}$ were grown by means of chemical vapor transport (CVT), which is the same method as that used in the crystal growth of CrNb$_3$S$_6$ \cite{Miyadai1983, Kousaka2009}.  The polycrystalline powders were synthesized as a precursor for the crystal growth by using gas phase method with a mixture of Mn, Ta and S in the molar ratio of $x_{\rm{nominal}}$ : 3 : 6.  The CVT growth was performed with an evacuated silica tube in a temperature gradient using iodine as a transport agent.  The precursor was placed at one end of the silica tube and then heated in the electric tube furnace under the fixed temperature gradient from \SI{1100}{\degreeCelsius} to \SI{1000}{\degreeCelsius} for 10 days. The single crystals were grown at the other end of the silica tube.
The obtained crystals have the shape of a hexagonal plane of 1 to 4 mm in diameter and of 0.1 to 0.5 mm in thickness.

Note that an emergence of chiral helimagnetism in the grown TMD crystals is very sensitive to $x_{\rm{nominal}}$ of the polycrystalline precursor.  It was reported that, in the case of CrNb$_3$S$_6$~\cite{Kousaka2022} and CrTa$_3$S$_6$~\cite{Mizutani2023}, the amount of Cr in the grown crystals was not necessarily fixed to a stoichiometric composition (the unity). Importantly, it was analyzed to be smaller than $x_{\rm{nominal}}$ of the polycrystalline precursors used in the CVT growth. As a consequence, $x_{\rm nominal}$ was tuned to be a value larger than the unity  in order to obtain single crystals without Cr defects that host the CHM and CSL.
These results indicate the importance of an optimization of the crystal growth condition in terms of $x_{\rm nominal}$.  Indeed, in the crystal growth of MnTa$_3$S$_6$, ferromagnetic single crystals were obtained from the precursors with the stoichiometric composition ($x_{\rm nominal} = 1$)~\cite{Zhang2018,Algarni2023}.

To examine the magnetic structure, zero-field SANS experiments were performed at BL-15 (TAIKAN) \cite{TAIKAN2015} at Material and Life Science Facility (MLF) in J-PARC, Japan.
The crystal was aligned with the $[001]^*$ direction parallel to the scattering vector, the incident neutron beam with a wavelength from 0.8 to 1.5 nm was directed along the $[1\bar{2}0]^*$ direction.  Scans in $\omega$ were performed to detect and maximize the magnetic satellite peak intensity at the (00$\delta$) and (00$\bar{\delta}$).
To obtain the magnetic scattering signals, large background intensity due to the incident neutron beam at low wave vector $Q$ region was subtracted from the SANS patterns at each temperature $T$.

The magnetization of the obtained single crystal was collected with $H$ applied in the direction perpendicular to the $c$-axis using a SQUID magnetometer (Quantum Design MPMS3).
The magnetization was measured as function of $T$ in the zero-field cooling (ZFC) and field-cooling (FC) processes. The magnetization was also examined in cycling $H$ at different temperatures.

\section{Results}

\subsection{Crystal optimization}
Single crystals were grown from the precursors with $x_{\rm nominal}$ from 1.00 to 1.35 for the $x_{\rm{nominal}}$ optimization.
As a result, the single crystal with $x_{\rm{nominal}}$ fixed to be 1.14 exhibited downward convex behavior of the magnetization,
which could be regarded as a manifestation of the CSL formation as discussed in other TMDs hosting the CHM and CSL \cite{Kousaka2009,Zhang2021,Obeysekera2021,Mizutani2023},  while the crystals from other $x_{\rm nominal}$ values showed ferromagnetic behavior.

\begin{figure}[tb]
\begin{center}
\includegraphics[width=8.5cm]{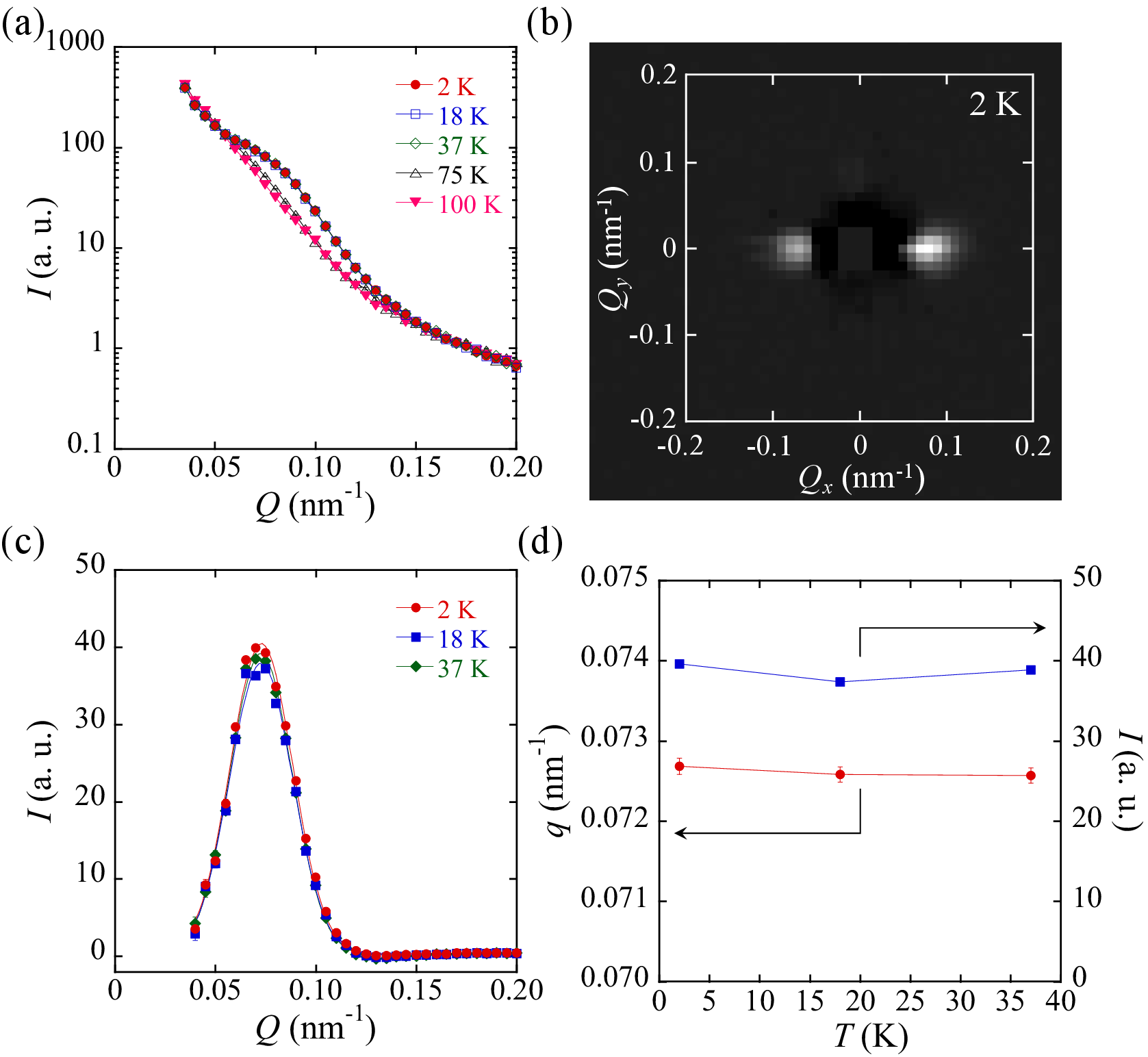}
\end{center}
\caption{
(a) The SANS profiles of the MnTa$_{3}$S$_{6}$ single crystal obtained by $Q$ scan along the $c$-axis measured at different temperatures.
(b) A map of the magnetic SANS intensity collected at 2 K.  The paramagnetic signal at 100 K was subtracted as a background signal from the original data.
(c) The magnetic profiles along the $c$-axis, processed by a subtraction of the incident neutron beam intensity as a background signal.
(d) A temperature dependence of the magnetic satellite peak position and peak intensity.
}
\label{f-SANS}
\end{figure}

\subsection{Small-angle neutron scattering (SANS)}

The SANS experiments revealed the presence of the CHM along the $c$-axis at zero $H$, as shown in Fig.~\ref{f-SANS}(a) which shows the $Q$-scan profiles along the $c$-axis at different temperatures.
Here, the SANS patterns were converted to the $Q$ profiles along the $[001]^*$ direction integrating in the signal over a 20 degrees angular sector.

As shown in Fig.~\ref{f-SANS}(b), a pair of the magnetic satellite peaks were clearly observed along the $[001]^*$ direction, which is parallel to $Q_{x}$, in the SANS pattern at \SI{2}{\kelvin} after the background subtraction using the paramagnetic data at 100 K.
As shown in Fig.~\ref{f-SANS}(c), the magnetic peak was clearly observed at \SI{0.073}{\nano \metre}$^{-1}$ in the $Q$-scan profiles along the $[001]^*$ direction after a background subtraction of the incident neutron beam intensity.  The peak position in $Q$-space produces a period of \SI{86}{\nano \metre} in real space. The magnetic period in MnTa$_{3}$S$_{6}$ is longer than those in CrNb$_3$S$_6$~\cite{Miyadai1983, Togawa2012} and in CrTa$_3$S$_6$~\cite{Kousaka2016}.
Moreover, additional SANS peaks were not detected at higher harmonic positions.
Therefore, these magnetic satellite peaks are the first real evidence of the presence of a CHM phase stabilized by the DM interaction in MnTa$_{3}$S$_{6}$ crystal.

Figure~\ref{f-SANS}(d) shows that the $Q$-position (the period of the helix) and intensity of the magnetic satellite peaks do not depend on the $T$ below a critical temperature $T_{\rm{c}}$ at 35 K, determined by magnetization measurements discussed later.
Such a tendency was found even in the $Q$ profile at \SI{37}{\kelvin},
which is slightly higher than $T_{\rm{c}}$.  The intensity drastically decreases with further increasing $T$ and vanishes in the $T$ regime for the paramagnetic phase,
as seen in the SANS data at \SI{75}{\K} and \SI{100}{\K} in Fig.~\ref{f-SANS}(a).

\subsection{Magnetic measurements}

\begin{figure}[tb]
\begin{center}
\includegraphics[width=8.5cm]{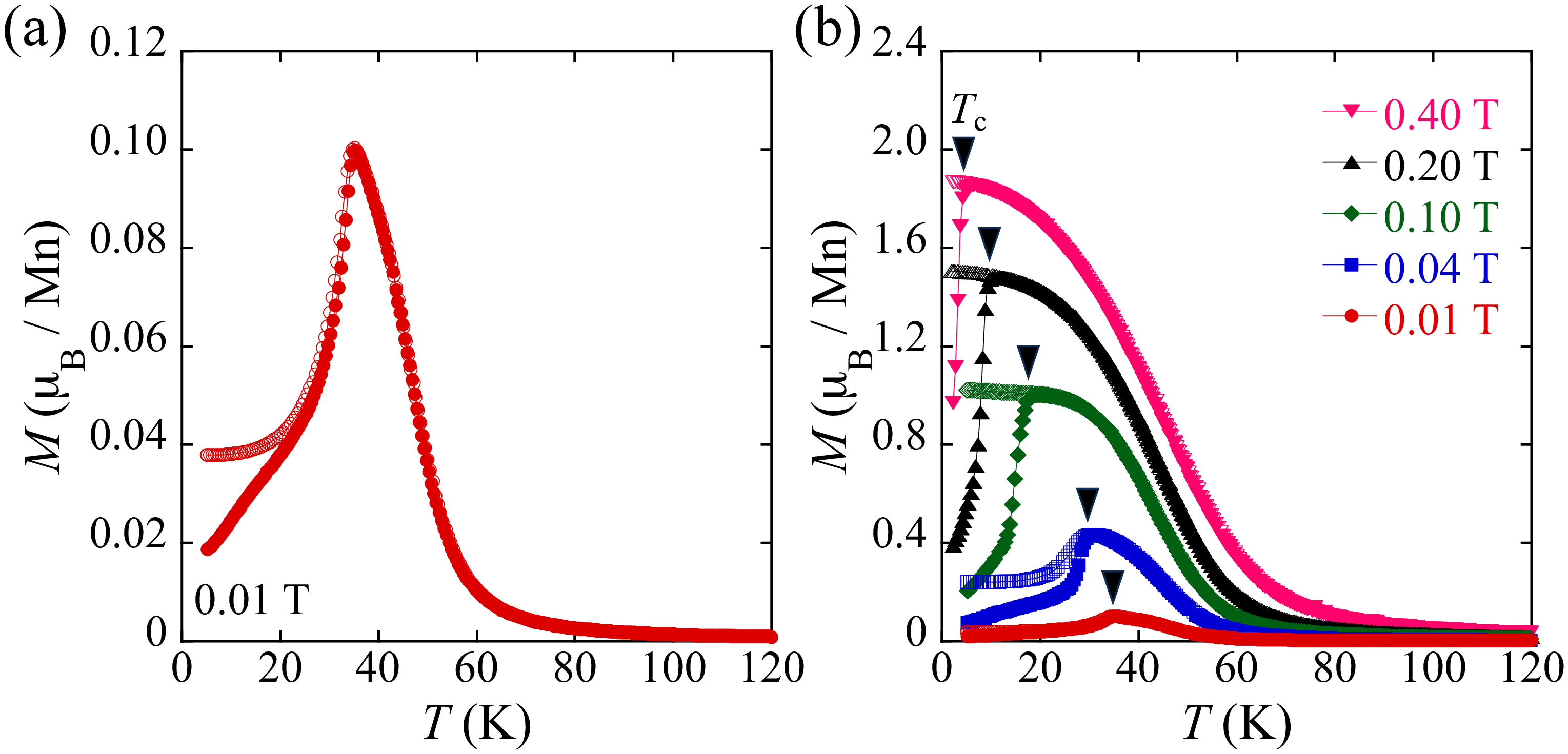}
\end{center}
\caption{
Temperature dependence of the magnetization in the MnTa$_{3}$S$_{6}$ single crystal measured with $H$ perpendicular to the $c-$axis collected at 0.01 T (a) and higher $H$s up to 0.40 T (b).
Closed and open marks denote the magnetization data collected in the zero-field cooling (ZFC) and field cooling (FC) procedures, respectively.}
\label{f-MT}
\end{figure}

The $T$ dependence of the magnetization at fixed magnetic field $H$ (\SI{0.01}{\tesla}) in  MnTa$_{3}$S$_{6}$ single crystal is shown in Fig.~\ref{f-MT}(a).
The Curie-Weiss temperature $\theta_{\rm CW}$, determined by fitting the inverse susceptibility in the paramagnetic region to a Curie-Weiss law, is 67 K.  
With decreasing $T$, the magnetization increases at around \SI{100}{\kelvin}.
After a rapid increase of the magnetization below around $\theta_{\rm CW}$, it sharply drops at \SI{35}{\kelvin}.
Here, a critical temperature $T_{\rm{c}}$ is defined at the peak position of the magnetization~\cite{Miyadai1983, Kishine2005}, as shown in Fig.~\ref{f-MT}(a).    With increasing the $H$ strength, the magnetization peak shifts toward lower $T$, as observed in Fig.~\ref{f-MT}(b).
Such behavior of the magnetization is consistent with those observed in CrNb$_3$S$_6$~\cite{Kousaka2009, Togawa2013, Kousaka2022} and CrTa$_3$S$_6$~\cite{Zhang2021,Obeysekera2021,Mizutani2023}.
In this respect, these results support the formation of CHM and CSL in the present MnTa$_3$S$_6$ crystal with the $H$ oriented perpendicular to the $c$-axis. 

Note that a deviation of the magnetization between the ZFC and FC procedures becomes evident in the $T$ regime below $T_{\rm{c}}$.
Figure~\ref{f-MT}(b) shows that, in the ZFC process, the magnetization peak survives till temperatures down to \SI{2}{\kelvin}, which corresponds to the lowest $T$ measured in this study,
while it disappears for $H$s at around \SI{0.1}{\tesla} in the FC process.

\begin{figure}[tb]
 \begin{center}
  \includegraphics[width=8.5cm]{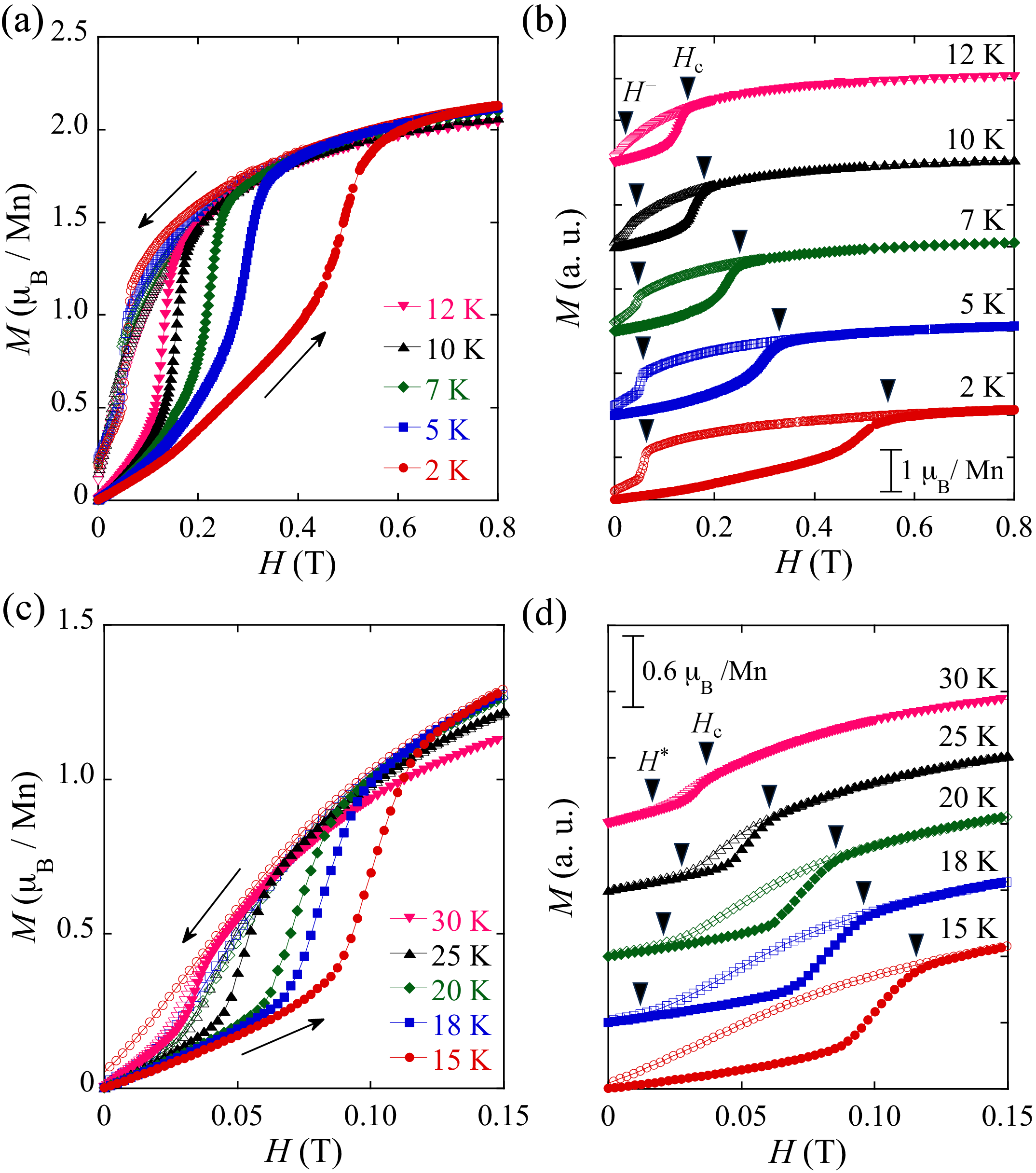}
  \caption{
Magnetic field dependence of the magnetization in the MnTa$_{3}$S$_{6}$ single crystal with $H$ applied in the direction perpendicular to the $c$-axis. The data were collected at temperatures from \SI{2}{} to \SI{12}{\kelvin} (a, b) and from \SI{15}{} to \SI{30}{\kelvin} (c, d).
Closed and open marks denote the magnetization data collected in the $H$ increase and decrease processes, respectively.
}
  \label{f-MH}
 \end{center}
\end{figure}
To examine the CSL formation in MnTa$_{3}$S$_{6}$, isothermal magnetization versus field curves were measured below $T_{\rm{c}}$. Note that the initial state was prepared via the ZFC process in every measurement in Fig.~\ref{f-MH}.
Downward convex behavior of the magnetization curves was observed in the $H$ increase process at temperatures ranging from \SI{2}{\K} to \SI{33}{\K}, which is consistent with the magnetization behavior expected for the CSL formation, as established in CrNb$_3$S$_6$~\cite{Kousaka2009, Togawa2013} and CrTa$_3$S$_6$~\cite{Zhang2021, Obeysekera2021}. This observation indicates that the CSL formation occurs in the present MnTa$_{3}$S$_{6}$ crystal.

Hysteresis behavior appears in the decreasing $H$ process ($0\le H_{\rm +max}\to 0$).
Here, we defined a critical magnetic field $H_{\rm c}$ as the irreversibility onset of the magnetization. The $H_{\rm c}$ value decreases with increasing $T$, as seen in Figs.~\ref{f-MH}(b) and~\ref{f-MH}(d).

In addition, a sharp drop of the magnetization was found at small $H$s when decreasing the $H$ in the low $T$ regime, as indicated by $H^{-}$ in Fig.~\ref{f-MH}(b).
It could be related with the penetration of chiral solitons into the system because of the disappearance of the surface barrier, as discussed in CrNb$_3$S$_6$~\cite{Shinozaki2016} and CrTa$_3$S$_6$~\cite{Mizutani2023}.
As a result, a distorted CSL with many dislocations should be formed at $H^{-}$, as experimentally observed in CrNb$_3$S$_6$~\cite{Paterson2019, Goncalves2020,Li2023}.
The $H^{-}$ position shifts toward smaller $H$s with increasing the $T$ and eventually it disappears at \SI{15}{\K}.

Alternatively, the magnetization gradually decreases in the decreasing $H$ process in the high $T$ regime, as seen in Fig.~\ref{f-MH}(d).
The magnetization curves for increasing and decreasing the $H$ merge, providing $H^{*}$ as the end point of the hysteresis.
Namely, the reversibility of the magnetization is recovered below $H^{*}$.
The behavior of $H^{*}$ with respect the $T$ shows that $H^{*}$ reaches the maximum value at \SI{25}{\K} and decreases at higher $T$ because of the suppression of $H_{\rm c}$.
In this regime, the net magnetization vanishes on average at zero $H$, which is an strong indication of the presence of the ideal CHM without any dislocation above \SI{15}{\K}~\cite{Paterson2019, Goncalves2020}.

\begin{figure}[tb]
\begin{center}
\includegraphics[width=8.5cm]{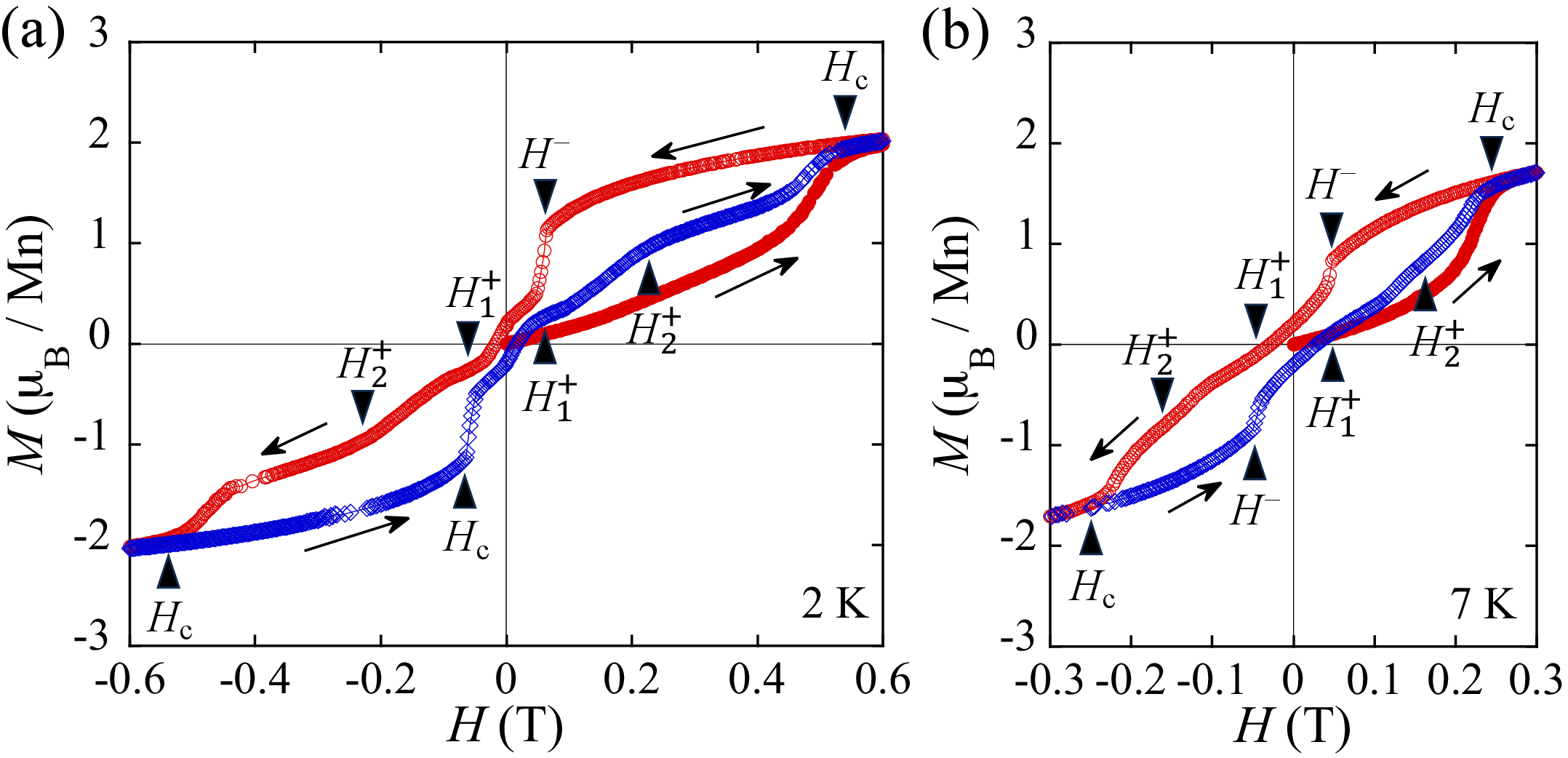}
\end{center}
\caption{A hysteresis loop of the magnetization in the MnTa$_{3}$S$_{6}$ single crystal at \SI{2}{\kelvin} (a) and \SI{7}{\kelvin} (b) with $H$ applied in the direction perpendicular to the $c$-axis.
The data, collected between \SI{1}{\tesla} and \SI{-1}{\tesla}, were zoomed to see the magnetization curves below $H_{\rm{c}}$.
Closed red circles denote the magnetization curve when increasing the $H$ from zero $H$ with the initial state prepared via the ZFC process.
Open red circles and blue diamonds denote the magnetization curves collected when decreasing and increasing $H$ processes, respectively.
}
\label{f-MHFull}
\end{figure}

Moreover, a residual magnetization remains at zero $H$ at the low $T$ regime, as seen in the $H$ decreasing curves in Fig.~\ref{f-MH}(a).
Such a spontaneous magnetization almost disappears at \SI{15}{\K}.

A hysteresis full loop of the magnetization was measured below \SI{15}{\K} at \SI{2}{\K} and \SI{7}{\K} as shown in Fig.~\ref{f-MHFull}.
It shows that, when decreasing, $0\le H_{\rm +max}\to 0$, (increasing, $0 \ge H_{\rm -max}\to 0$) the $H$ to zero after being saturated at positive (negative) large values,
$H_{\rm +max}$ ($H_{\rm -max}$), and passing through $H^{-}$, upward convex behavior of the magnetization appears in the vicinity of zero $H$.

This feature could be explained due to a reorientation process of the magnetic moments via their collective rotation so as to recover the CHM.
Then, downward convex behavior, which could be associated with the CSL formation, starts to appear above a given $H$.
Such behavior was observed repeatedly at the low $T$ regime, as indicated by the onsets of $H^{+}_{1}$ and $H^{+}_{2}$ in Figs.~\ref{f-MHFull}(a) and~\ref{f-MHFull}(b).

\begin{figure}[tb]
\begin{center}
\includegraphics[width=8.5cm]{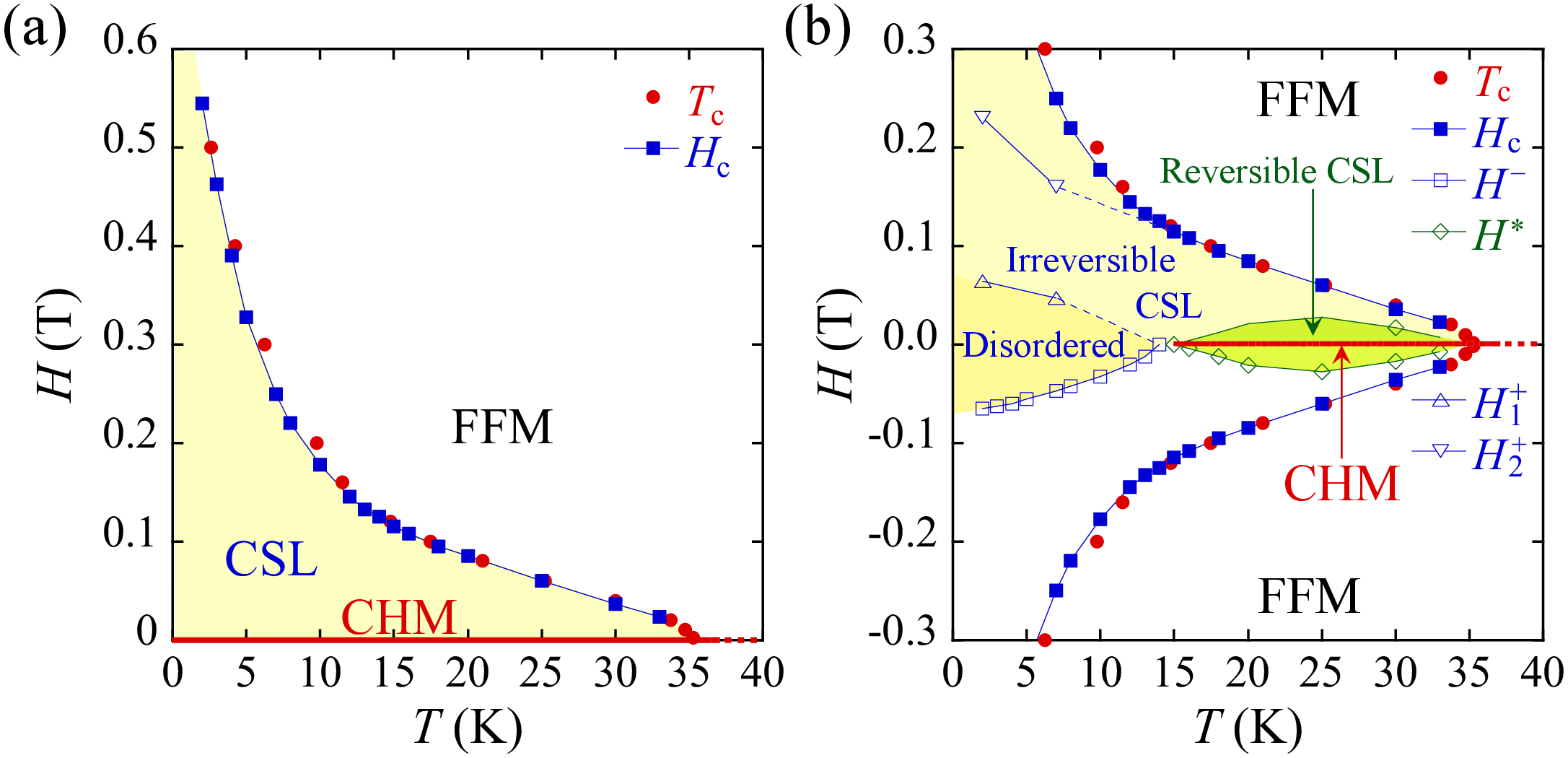}
\end{center}
\caption{Magnetic phase diagram of the MnTa$_{3}$S$_{6}$ single crystal, determined by the magnetization curves.
The data were collected with the initial state in ZFC process (a) and taken by hysteresis loops with the $H$ sweep from $-1$ T to 1 T (b).
}
\label{f-PhaseDiagram}
\end{figure}

The experimental data obtained in the present MnTa$_{3}$S$_{6}$ crystal are summarized in the phase diagram depicted in Fig.~\ref{f-PhaseDiagram}.
A simple phase diagram is obtained when measuring using the ZFC protocol, as shown in Fig.~\ref{f-PhaseDiagram}(a).
The CHM exists at zero $H$ below $T_{\rm c}$, as confirmed in the SANS data, and it transforms into the CSL with increasing the $H$ applied perpendicularly to the $c$-axis.
$T_{\rm c}$ and $H_{\rm c}$ coincide with each other, providing the phase boundary between the CSL and forced ferromagnetic (FFM) state.  Interestingly, the boundary line expands toward higher $H$s below \SI{15}{\K}.
An extrapolation of the $H_{\rm c}$ boundary line below \SI{15}{\K} gives \SI{0.7}{\tesla} at \SI{0}{\K}, while an extrapolation of the boundary at high temperature regime provides \SI{0.2}{\tesla}.
The origin of this feature remains to be clarified although similar behavior was reported in MnNb$_{3}$S$_{6}$~\cite{Ohkuma2022}.

Figure~\ref{f-PhaseDiagram}(b) shows the phase diagram when decreasing the $H$ from the saturated FFM state to zero ($0\le H_{\rm +max}\to 0$).
The hysteresis survives far below $H_{\rm c}$, indicating an expansion of the FFM region. The remanent state appears at zero $H$ below \SI{15}{\K}.
This magnetic state should correspond to the distorted CHM since it is generated from the disordered CSL with soliton defects that is formed at $H^{-}$~\cite{Paterson2019, Goncalves2020,Li2023} and leaves the residual magnetization even at zero $H$.
Moreover, the CSL region is likely to be divided into the low and high $H$ regime below \SI{15}{\K}, as recognized by the magnetization anomaly (given by $H^{+}_{1}$ and $H^{+}_{2}$) in the full loop of the magnetization.

On the other hand, a doom-shaped region, enclosed by $H^{*}$, appears in the high $T$ regime between \SI{15}{\K} and $T_{\rm c}$.
Here, the CHM appears at zero $H$ and transforms into the CSL when increasing the $H$.
Nonlinearity in the CSL develops slowly in this regime~\cite{Kishine2015, Togawa2016}, although irreversible behavior of the magnetization becomes evident above $H^{*}$. 

The magnetization curve at \SI{15}{\K} is quite interesting since the gradual reduction of the magnetization continues down to zero $H$ with decreasing the $H$ ($0\le H_{\rm +max}\to 0$) and is followed by the CSL formation when increasing again the $H$.
Further investigations are required to reveal such a continuous magnetization process~\cite{Mito2018} observed at \SI{15}{\K} in Fig.~\ref{f-MH}(d).

In summary, our experiments show unambiguously that chiral helimagnetism is present in MnTa$_3$S$_6$ single crystals when they are carefully grown.
The CHM phase was directly observed as a pair of satellite peaks in the SANS experiments.  The magnetization measurements support the presence of CHM and CSL phases as well.
However, the irreversible behavior observed in the magnetization curves in MnTa$_3$S$_6$ need to be still clarified.
We determined the phase diagram of MnTa$_3$S$_6$, which is dependent on the initial state at zero $H$, in contrast with those observed
in other TMDs such as CrNb$_3$S$_6$ and CrTa$_3$S$_6$.
The material parameters of TMDs hosting the CHM and CSL are summarized in Table~\ref{tbl-params}.
Detailed analyses using SANS under applied $H$s, resonant  X-ray magnetic scattering~\cite{Honda2020} and Lorentz microscopy~\cite{Togawa2012, Paterson2019, Li2023} are required for further understanding of the nature of different magnetic states observed in the proposed magnetic phase diagram.

For instance, the temperature dependence of the magnetic satellite intensity should be examined in detail using SANS technique, which will allow an evaluation of the $T$ evolution of the DM interaction constant.
Furthermore, the presence of the remanent state in MnTa$_3$S$_6$ is worth investigation since it gives coercivity to chiral solitons in the CSL system.
Namely, tunable capability of the soliton number at zero $H$ emerges in TMDs, which may open up the possibility to develop memory bits using the CSL system. 

\section*{Acknowledgments}

This work was supported by JSPS KAKENHI Grant Numbers 17H02815, 19H05822, 19H05826, 19KK0070, 21H01032, 23H01870 and 23H00091.
Grants No. PID2022-138492NB-I00-XM4 funded by MCIN/AEI/10.13039/501100011033, OTR02223-SpINS from CSIC/MCIN and DGA/M4 from Diputaci\'on General de Arag\'on (Spain) are also acknowledged.
The SANS experiments at the MLF of the J-PARC were performed under user programs (Proposals No. 2020C0001, 2021B0265, 2022B0232 and No. 2023A0167).

\section*{Data Availability Statement}
The data that support the findings of this study are available from the corresponding author upon reasonable request.

\end{document}